# Feature Specification and Refinement with State Transition Diagrams [*]

*Cornel Klein, Christian Prehofer, Bernhard Rumpe*
*Institut für Informatik, Technische Universität München*
*80290 München, Germany*
*e-mail: (klein|prehofer|rumpe)@informatik.tu-muenchen.de*

**Abstract**

In this paper, we introduce a graphic specification technique, called state transition diagrams (STD), and show the application to the feature interaction problem. Using a stream-based formal semantics, we provide refinement rules for STDs. Refinements define an implementation relation on STD specifications. We view features as particular refinements which add previously unspecified behavior to a given STD specification. The refinement relation is then used to add features, and to define the notion of conflicting features. Our techniques are demonstrated by a systematic development of an example given in [25].

**Keywords:** state transition diagrams, specification, refinements, feature interaction, formal methods, automata

## 1 Introduction

In the last couple of years, the problem of defining and implementing interacting services, also called *features* has received considerable attention [11, 24]. The problem of such interactions is that some feature has to behave differently in the presence of the other, due to an interaction. In the literature, many *interaction* detection and resolving mechanisms for *feature interactions* have been proposed [4, 12].

Many techniques in the literature describe features only at an implementation-oriented level. In this paper, we introduce state-transition-diagrams (STDs) which are capable of defining component properties at different abstraction levels. We use such STDs for the abstract specification of a complete component, followed by provably correct refinements to a more implementation oriented level. Refinements define an implementation relation on STD specifications. Thus we can view features as particular refinements which add previously unspecified behavior. The refinement relation is used to incrementally add features to an existing system, as well as to define the notion of conflicting features.

A specification defines properties on which the environment of a component can rely on, whereas an implementation is more detailed, e.g. by efficiency considerations. Thus a specification is usually much more *underspecified* than the implementation is. This allows to consider the abstract component description from the environment point of view. For this purpose, we describe features by very liberal specifications, which only specify the desired behavior. Other cases are fully unspecified and are semantically modeled by non-determinism. This non-determinism is later reduced via refinements.

---

[*]This paper partly originated in the SYSLAB project, which is supported in part by the DFG under the Leibnizpreis and by the Siemens-Nixdorf corporation.

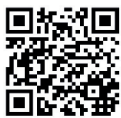

With our techniques for under-specification we can define feature interactions in terms of refinement. We simply call two features *conflicting* if there is no common refinement or implementation.

It is important to stress that our techniques have been developed in the context of distributed, reactive systems, for which telecommunication systems are typical examples [9, 7]. A component in a reactive system continuously interacts with its environment, which consists of other components, by the exchange of messages. However, for the exposition here, we do not elaborate concurrency issues and instead focus on features as refinement.

Many other models and notations for specifying behavior have been developed. Examples for graphical notations are SDL process graphs [5], statecharts [15], state transition diagrams ([14], [23]). Other, formally based approaches are I/O-automata [20] or TLA [19]. We claim that our model combines the advantages of a graphical notation with elaborated and formal refinement rules.

The paper is structured as follows. In the following section, we briefly introduce our system model. Based on this semantic model, we present state transition diagrams and refinement rules in Sections 3 and 4. Features and their interaction are discussed in Section 5. In Section 6, we show how state transition diagrams and the refinement calculus can be used for the incremental addition of features to an existing system. In Sections 7 and 8, we discuss related work and draw some conclusions.

## 2   The Semantic Model

The semantic model we are using is based on *stream processing functions*. It is in detail presented in [10, 9, 18, 13] and it is based on dataflow networks as originating from the work of G. Kahn [16]. We model a component of a system as an entity communicating asynchronously with its environment by the exchange of messages. We restrict components to have exactly one port where messages arrive and one port where messages are sent. The input and output message sets $I$ resp. $O$ constitute the *syntactic interface* of the component. Formally we define $\Sigma = (I, O)$ as the signature of a component.

**Example:** Consider a simple telephone line. The set of input messages $I_{tel} = \{LT, OH\} \cup \{DL(n)|n \in Nums\}$ contains the messages *LT* ("Lift receiver"), *OH* ("Receiver on hook") and $DL(n)$ ("Dial number $n$"). The set of output messages $O = \{RG, CT, DT, BY, HU\}$ contains the messages *RG* ("Phone Rings"), *CT* ("Connection established"), *DT* ("Dial Tone starts"), *BY* ("Callee Busy") and *HU* ("Callee Hang Up").

Let for some set $M$ the set of all finite sequences over $M$ be denoted by $M^*$ and the set of all infinite sequences over $M$ be denoted by $M^\infty$. The set of *streams* over $M$ is defined by $M^\omega = M^* \cup M^\infty$. With $\epsilon$ we denote the empty sequence, with $[m_1, \ldots m_n]$ we denote the finite sequence of messages $m_1 \ldots m_n$, and with $l : k$ the concatenation of $l$ and $k$.

In order to specify components, we have to specify the possible observations of the behavior of a component from the viewpoint of the environment. Given a component with signature $\Sigma = (I, O)$, the environment of the component is able to send an input stream from $I^\omega$ to the component, and to observe a message stream from $O^\omega$ produced by the component. To set both in causal relation, we use the concept of a *stream processing function*. A stream processing function $f$ relates a stream $i \in I^\omega$ of incoming messages to the reaction, a stream of $f(i) \in O^\omega$ of outgoing messages. A stream processing function is total, which formally models the fact that a component in an asynchronous system model has to accept all incoming messages in every possible order and cannot reject any of them.

In order for a stream processing function to be an adequate model of a component in an information processing system, we have to make sure that a component can not predict the future. In particular, a stream processing function in our system model may only enlarge its output stream if more input arrives, but it may not remove messages from the output stream that it has produced previously. This is formally captured by the following *monotonicity requirement*:

$$\forall i, j \in I^\omega . i \sqsubseteq j \Rightarrow f(i) \sqsubseteq f(j)$$

Here, $\sqsubseteq$ denotes the prefix relation between streams, i.e. $i \sqsubseteq j$ holds iff there exists a $k$ such that $i : k = j$. In the sequel, by $I^\omega \rightarrow O^\omega$ we denote the set of all stream processing functions satisfying the above monotonicity requirement. For a more comprehensive treatment of stream processing functions, see [9] and [10]. In these papers, also an extension of the model for real-time is presented.

While *one* stream processing function can be used to model a *deterministic* component, we also have to take into account *nondeterminism* and *under-specification*. Nondeterminism occurs during runtime of a component due to non-deterministic choice inside a component. Under-specification is used during the development process, where a component is refined by gradually introducing more and more requirements. Since from the point of view of the environment it does not matter whether choices are made at development-time or at run-time, both concepts coincide. We therefore use *sets of stream processing functions* to model under-specification and nondeterminism of the behavior of a component. The use of sets of stream processing functions in our model is a necessary prerequisite for the refinement calculus as defined in Section 4.

**Example (continued):** If on a telephone line, the receiver is lifted from a phone, and a number is dialed, the user gets a dial tone (*DT*), followed by either a ring-signal (*RG*), or a busy-signal (*BY*). Call the set of stream processing functions modeling the telephone line *TEL*. This requirement can be specified by the following formula:

$$\forall f : f \in TEL \Rightarrow f([LT, DL]) = [DT, RG] \vee f([LT, DL]) = [DT, BY]$$

## 3  State Transition Diagrams

In order to enhance the acceptance of formal approaches for software specification and development, graphical and textual notations are needed, which provide a convenient "user interface" to formal models for software systems. Therefore, in this section we introduce a graphical description technique - *state transition diagrams* - along the lines of [14, 23, 22] for the specification of components in our semantic model.

State transition diagrams are based on the concept of a *state machine* (STM), which we introduce first. A state machine describes the behavior of a component using states of the component and transitions between states. There are two kinds of transitions, *external transitions* and *internal transitions*:

*External transitions* are labeled by an input message and a sequence of output messages. The transition is enabled if its input message ("stimulus") has been received by the component. *Internal transitions*, however, are labeled with an empty input sequence and a sequence of output messages. They are enabled without the occurrence of an external stimulus.

Operationally, if a transition is enabled, it may be taken, in which case the component emits the output message of the transition and changes its internal states. If several transitions are enabled in a certain state, one of them is chosen nondeterministically. As already pointed out in Section 2, this allows to model nondeterminism as well as under-specification. It also is an absolutely necessary prerequisite for the refinement calculus for state machines.

Formally, if $I^\epsilon = I \cup \{\epsilon\}$ denotes the set of input messages $I$ extended by the empty message sequence $\epsilon$, a *state machine* is a tuple $STM = (S, I, O, \delta, S^0)$, consisting of:

- a nonempty set of states $S$,
- a nonempty set of input messages $I$,
- a nonempty set of output messages $O$,
- a transition relation $\delta \subseteq S \times I^\epsilon \times S \times O^*$, and
- a nonempty set of initial states $S^0 \subseteq S$.

The important point is how we deal with a partial transition relation, i.e. with states $s$ and input $i$ for which $s'$ and $o$ exist such that $\delta(s, i, s', o)$ holds. In this case, we assume that the behavior of the component is completely unspecified, i.e. the component may expose an arbitrary behavior. Refinement (see Section 4) allows to incrementally specify these cases, which are ignored at first place.

Note that none of the above given sets needs to be finite. A finite representation is obtained by a graphical notation, called *state transition diagrams* (STDs) [14]. For brevity, we do not introduce STDs formally, but by the use of an example:

An *interactive stack* is a component storing a stack of integers. The stack can only be accessed by sending messages to it. The message $Push(a)$ requests the stack to push the integer $a$ on the top of the stack. The message $Pop$ requests the stack to throw away its top element and the message $Top$ requests the stack to deliver its top element. A STD specifying the stack looks as follows:

```
std stack = {
    input     Push(Int)| Pop | Top
    output    Int
    attributes l :: [Int]
```

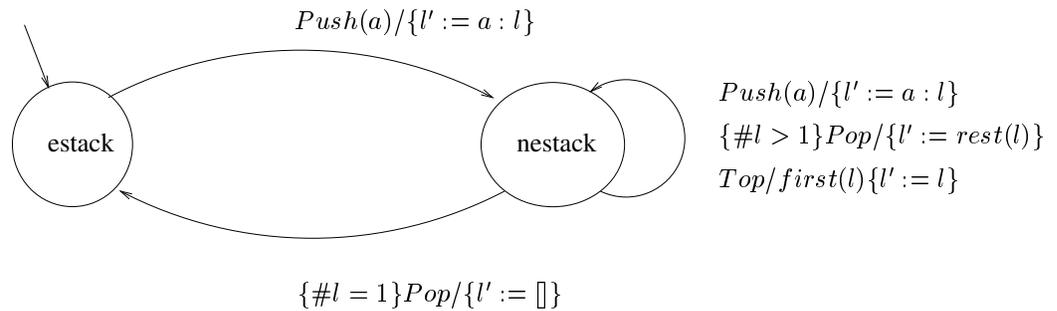

}

This specification can be explained as follows:

- Lines 2 and 3 specify the syntactic interface of the specified component, i.e. the components $I$ and $O$ of the state machine. For defining these sets, a notation similar to datatype declarations in functional languages like ML or Haskell is used.
- The state space of a component is given by a set of attributes, each having a name and a type. In our case, the stack has an attribute $l$ of type *[Int]*, where *[Int]* denotes the set of all sequences of integers.
- To specify the behavior of a component, we use a finite directed graph, consisting of a set of control states (or "vertices"), and a set of transitions (or "arrows"). In our example, the control states are named *estack* and *nestack*, representing the empty stack and the non-empty stack, respectively. Therefore, the state space $S$ of the state machine consists of a data part, i.e. the attributes, and of a control part, i.e. the set of nodes.

    Initially, in our example the automaton is in the start state *estack*, indicated by an arrow without source state.
- Transitions of the form $\{P\}m/op\{Q\}$ are labeled by a precondition $P$, a message (possibly with parameters), a postcondition $Q$ and an output expression $op$. Output expressions are arbitrary expressions which specify the output of a transition on the different ports. For pre- and postconditions, we allow arbitrary formulae.

    Intuitively, transitions in the STD are translated to transitions of the state machine as follows: A transition $\delta(s, i, s', o)$ occurs in the state machine, if
    - input $i$ matches with the input message of a transition of the STD,
    - input $i$ and state $s$ satisfy the precondition of the transition and
    - input $i$, state $s$, output $o$ and state $s'$ satisfy the postcondition of the transition.

## 4  Refinement of State Transition Diagrams

The semantics of a *STM* $= (S, I, O, \delta, S^0)$ is given as a set of stream processing functions (see Section 2). A stream processing function maps an input stream $i$ of incoming messages to a stream $f(i)$ of output messages. Thus a stream processing function is the black-box description of the behavior of a deterministic component. The use of a set of functions as semantics models the nondeterminism of components [14, 22].

State transition diagrams are propositions about the behavior of components in our semantic model. By refining a component, we can add more and more propositions about its behavior. The refinement relation on sets of stream processing functions is fairly simple: As each function describes one possible behavior, behavioral refinement is defined by *set inclusion* on sets of stream processing functions. This kind of refinement corresponds to logical implication, i.e. each behavior satisfying a refined component satisfies also the original component. We therefore define the *refinement relation* $\leadsto$ between STMs through the refinement relation between their semantics:

$$STM_1 \leadsto STM_2 \quad \text{iff} \quad [\![STM_2]\!] \subseteq [\![STM_1]\!]$$

An important point about this notion of refinement is that it is *compositional*, meaning that we can locally refine individual components without affecting the global behavior of a system [8].

However, the above semantical characterization is not yet useful from a practical point of view. Therefore, in [23, 22, 17] a refinement calculus working on STDs is given and proven correct. The refinement calculus is very tractable, since the applicability of the refinement rules can be checked almost automatically on the syntactical level. In particular, the following rules are sound:

1. *Addition of new states*. New states may be introduced, since they are unreachable and therefore do not affect the behavior.

2. *Removal of states*. Likewise, states which are not reachable may be removed.

3. *Refinement of states*. A state may be partitioned into several ones, if all transitions leaving or ending in the state are partitioned accordingly.

4. *Addition of transitions*. If in a state $s$ no transition exists for a certain input $i$, the behavior of the component $s$ for the input is completely unspecified. Therefore, an arbitrary set of transitions labeled with $s$ as source state and $i$ as input message can be added. Note, that if an internal transition with the same source state exists and the precondition does not evaluate to false, the internal transition can always be taken and therefore no addition of transitions is allowed.

5. *Removal of transitions*. If in a state several transitions with the same input message exists, then one of them is chosen nondeterministically or some $\epsilon$-transition is taken. Therefore, these transitions may be removed, as long as there exists at least one transition labeled with this input or with $\epsilon$.

6. *Removal of initial states*. Initial states may be removed, as long as at least one initial state remains.

These rules are the basis for the formal development in Section 6.

## 5 Features and their Interaction

A *feature* is defined as a refinement of an existing component by a sequence of refinement steps of the above refinement calculus. Usually, new states and new transitions are introduced. A feature can therefore be regarded as a proposition about the behavior of a component. Refinement ensures that the new component satisfies all propositions of the original component as well as the newly added feature.

Given an existing system $S$, we denote the refinement step introducing a new feature $F$ and leading to an enhanced system $S'$ by $S \stackrel{F}{\leadsto} S'$. Given an existing system $S$, we call two features $F'$ and $F''$ *conflicting*, if we may independently apply $S \stackrel{F'}{\leadsto} S'$ and $S \stackrel{F''}{\leadsto} S''$ but there exists no $\hat{S}$ such that

$$S' \stackrel{F''}{\leadsto} \hat{S},$$
$$S'' \stackrel{F'}{\leadsto} \hat{S}.$$

In other words, it is not possible to find a common refinement of $S$ incorporating both features. The problem of checking if a set of features $F_1, \ldots F_n$ can be integrated into an existing component $S$ can therefore be reduced to the problem of finding refinements such that $S \stackrel{F_1}{\leadsto} S_1 \ldots S_{n-1} \stackrel{F_n}{\leadsto} S_n$. Note that this refinement $S_n$ is still an abstract specification of the system behavior but it incorporates all features and contains the description how the features work together. This specification can be further refined in our refinement calculus, leading to deterministic state transition diagrams specifying the behavior in an implementation-oriented way.

| Name | Description | Data Structure |
|---|---|---|
| Forwarding Features | | |
| Follow Me (*FM*) | assign subscriber to a different phone | *FM*: $DN \rightarrow DN$ |
| Delegate (*DL*) | delegate call to other subscriber | *Del*: $DN \rightarrow DN$ |
| Delegate Busy (*DB*) | as above, if phone busy | *DelB*: $DN \rightarrow DN$ |
| Follow Me No Answer (*FMNA*) | assign subscriber to a different phone, if busy | *FMNA*: $DN \rightarrow DN$ |
| Delegate No Answer (*DLN*) | delegate call to other subscriber, if no answer | *DelNA*: $DN \rightarrow DN$ |
| Do Not Ring (*DNR*) | do not alert this telephone | *DNR*: $DN \rightarrow \mathbb{B}$ |
| Blocking Features | | |
| Vacation Protection (*VP*) | do not connect calls to this subscriber | *VP*: $DN \rightarrow DN$ |
| Originating Call Screening (*OCS*) | block incoming calls from certain *DN*s | *OCS*: $DN \times DN \rightarrow \mathbb{B}$ |
| Terminating Call Screening (*TCS*) | block outgoing calls to certain *DN*s | *TCS*: $DN \times DN \rightarrow \mathbb{B}$ |
| Calling Number Delivery Blocking (*CNDB*) | refuse call identification on outgoing call | *CNDB*: $DN \rightarrow \mathbb{B}$ |
| Anonymous Call Rejection (*ACR*) | block incoming calls without caller identification | *ACR*: $DN \rightarrow \mathbb{B}$ |

Figure 1: Telephone Features from [25]

Note that feature interactions as defined above occur only if two features $F'$ and $F''$ can be applied independently of each other. If one feature depends on the other, e.g. using a state introduced by the other, a conflict may not occur in the above sense. In this case the developer of the second feature has to be aware of the first one, such that he can resolve all possible conflicts.

The advantage of this specification approach is that it abstracts from irrelevant implementation details at first place. This allows to concentrate on the main issues of feature interaction, and considerably enhances the probability that two feature specifications have a common refinement.

To achieve a feature integration of conflicting features, some of the features have to be adapted. This however means that some specification is revised, and not refined. In this case, behavior guaranteed previously, and therefore assumed by the environment, is not ensured anymore. Therefore, the environment has to be adapted to this changed behavior as well.

## 6 Example: Call Processing

The section shows a nice example for call processing by an elegant model which was proposed by Zave in [25]. We re-develop this example strictly by refinement, i.e. services are added via refinements. Since refinement is an associative operation, we can modularly add several features to a given specification. In addition, we model the example in [25] by only one automaton. In contrast, this is described by Zave with a simple finite-state automata, some pseudo-code and several tables in [25]. Furthermore, "refinement" is not formalized but modeled by "replacing" some tables. (It is clearly a matter of taste to specify (parts of) the automaton as a table, but this is not pursed here.)

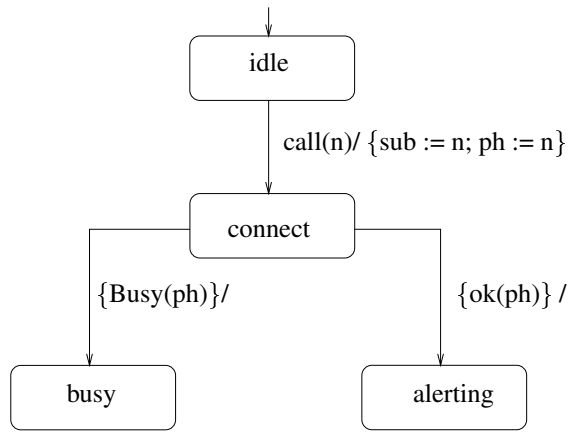

Figure 2: Basic Connect Model

The example models the process of connecting a call. We just model the core switching unit, abstracting from several other issues which are modeled as predicates. For instance, we assume a predicate "Busy" which indicates if a phone is busy.

As shown in [25], most typical feature interactions can be avoided by a conceptual distinction between a subscriber and a (physical) telephone (also named *DN* for directory number). Figure 1 shows the features and the used data-structures. For a detailed explanation and motivation we refer to [25]. The column "data structure" contains fixed (for each call) functions or predicates which determine the behavior of the feature. For instance, *FM* is a function which computes the new *DN* to which the call is forwarded. Similarly, predicates like *VP* and *OCS* determine if the feature is enabled for a *DN* or a pair of *DN*s, respectively. For more details and motivation, we refer to [25].

**A Systematic Development**

In the following, we develop a STD which models the features of Figure 1. We start with a rather simple model, which is refined in a stepwise way. We justify the refinements by the rules of the last section, but only give informal proofs.

For the following development, the underlying call processing model has to be quite general, as e.g. it has to encompass the separation of subscriber and phone. This is needed to enable the addition of features. In practice, this means that we may have to backtrack to earlier design stages, in order to accommodate for later steps. Although this method is quite rigid, it often enforces good design.

**Step 0:** The STD in Figure 2 shows a very abstract model of a switching unit. A call is invoked by a message "call" and finally ends either in state "busy" or "alerting" (i.e. ringing the phone). Note that we use variables "sub" and "ph" for subscriber and phone, respectively.

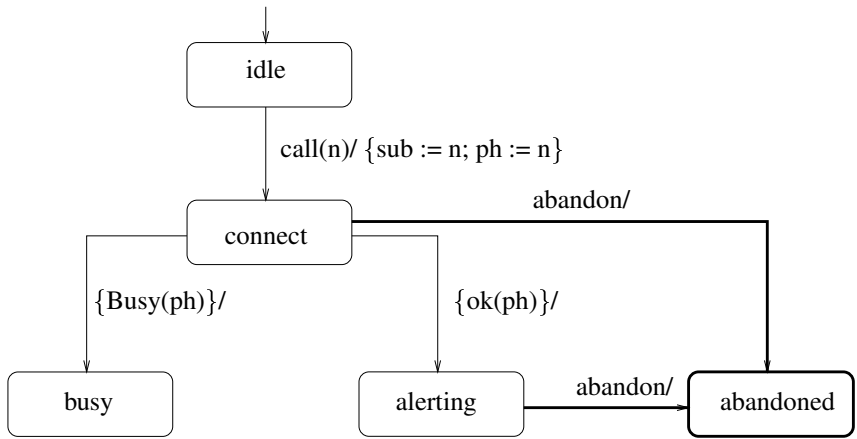

Figure 3: Adding State Abandoned (Step 1)

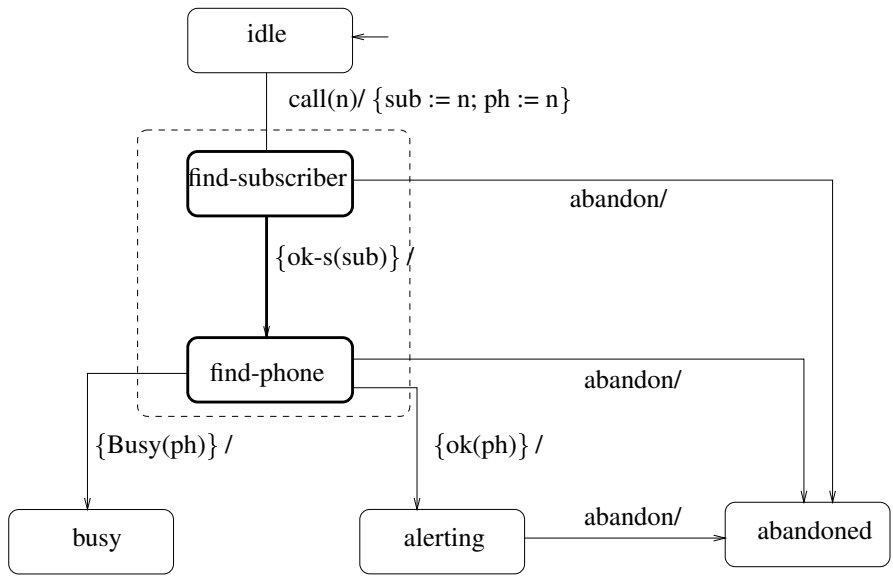

Figure 4: Splitting State Connect (Step 2)

We further use the predicates *Busy: DN → $\mathbb{B}$*, which indicates if a *DN* is busy, and *ok: DN → $\mathbb{B}$*, which denotes (yet underspecified) successful cases. For refinement, it will be important to assume that ok is disjoint from the predicates on all further added transitions. Note that we delay such design decisions via under-specification.

**Step 1:** The first refinement step adds a state abandoned, shown in Figure 3, where the newly added states and transitions are drawn with bold lines. This new state models the case of a hook-up. The newly added transitions are invoked by the new message "abandon". Note that this is a proper refinement step, since in Figure 2, the behavior of the STD wrt the message abandon is underspecified and fully unpredictable.

**Step 2:** In the second step, a non-trivial refinement step is performed, namely splitting the state connector into two states, "find-subscriber" and "find-phone". This is shown in

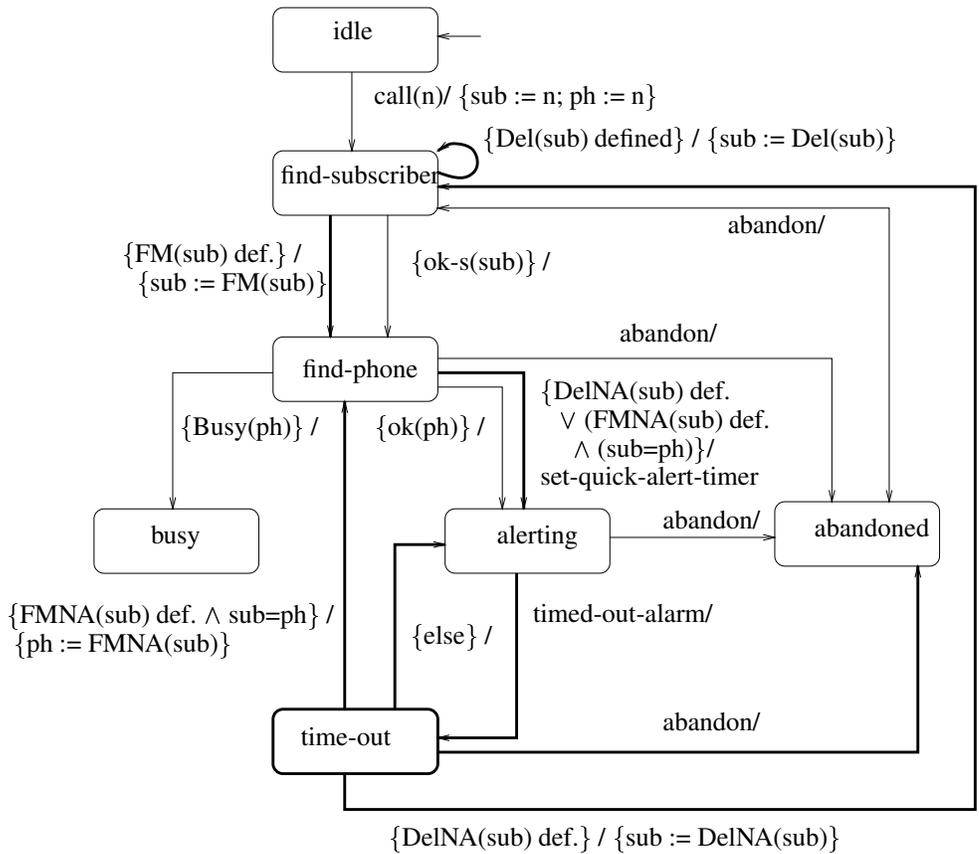

Figure 5: Adding Forwarding Features to Figure 4 (Step 3)

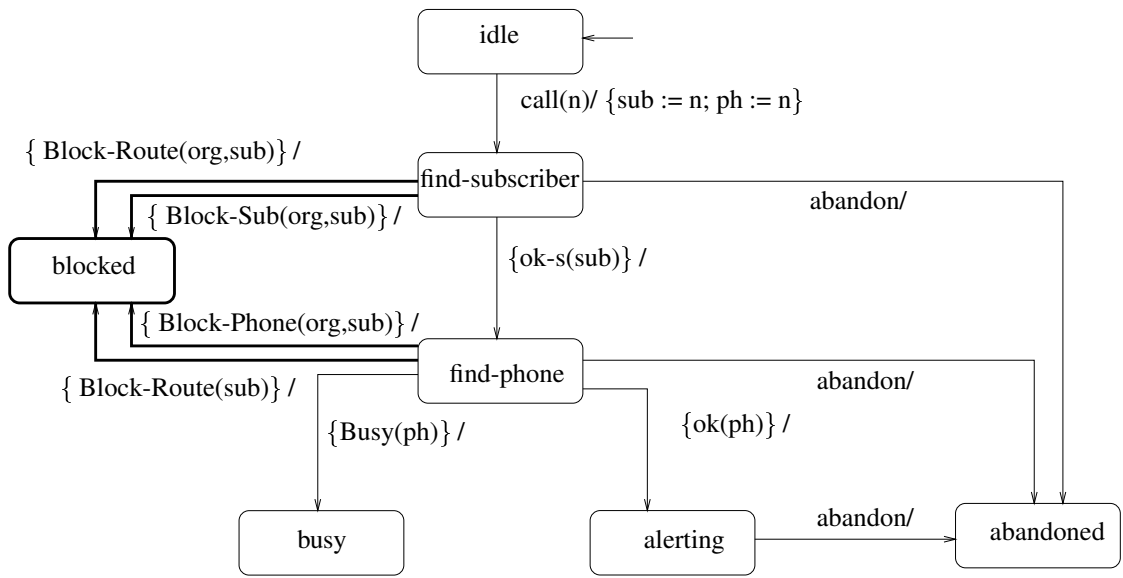

Figure 6: Adding Blocking Features to Figure 4 (Step 4)

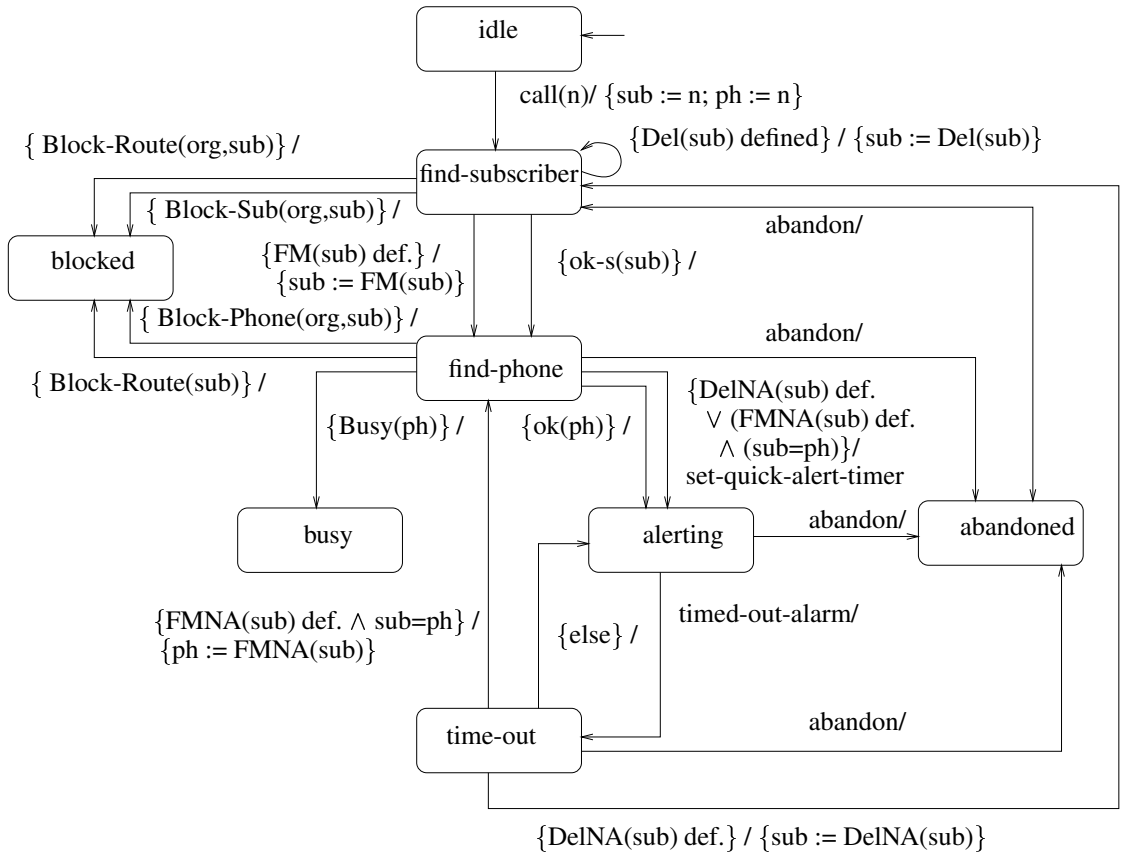

Figure 7: Integrating Forwarding- and Blocking Features (Step 5)

Figure 4. Note that the new transition between the new states has a condition with a new predicate "ok-s", which is again underspecified. For refinement, it is not strictly needed to add a transition from find-subscriber to abandoned, but we consider it useful here.

After generalizing the structure in the last step, we now add the features of Figure 1. Each of the two sets of features in Figure 1 is first added individually to the STD in Figure 4. Then we argue that they can be combined.

**Step 3:** Figure 5 shows additional transitions and a new state time-out for the forwarding features. The new state "time-out" is needed for the delegation-on-no-answer features. Note that the condition {*else*} is a shorthand for the negation of all other conditions. In the new transition to alerting, we assume an external timer, which is set by "set-quick-alert-timer". It is assumed to send the message "time-out-alarm", if the call is not answered.

To show that the added transitions give a proper refinement, we must assume that the ok-s condition on the transition from the state find-subscriber is disjoint from the new transitions. In other words, these transitions model cases not handled (and hence not specified) in the previous STDs. Hence we again make use of our earlier under-specification. Furthermore, we must assume that we do not introduce infinite loops with the new transitions, e.g. with the transition with condition "*Del(sub)* defined". This is another important assumption needed for refinement. The added state poses no problem wrt refinement, since the new state is only reachable by a new transition.

**Step 4:** Finally, figure 6 shows how the blocking features are added to the STD in Figure 4. We introduce a new state named "blocked". We also assume here an additional variable "org" for the origin of the call. The conditions in the new transition assume new predicates, which can be defined in terms of the data structures of these features as follows (see [25]):

$$\begin{aligned}
\textit{Block-Sub(origin,sub)} &= \textit{DNR(sub)} \lor \textit{CNDB(orign,sub)} \lor \textit{ACR(orign,sub)} \\
\textit{Block-Phone(origin,sub)} &= \textit{VP(sub)} \\
\textit{Block-Route(origin,sub)} &= \textit{OCS(origin,sub)} \lor \textit{TCS(origin,sub)}
\end{aligned}$$

For a proper refinement, we again assume that the transition with condition ok-s is disjoint from the new transitions, and hence again use our earlier under-specification.

**Step 5:** The last step, integrating the added features of Step 3 and Step 4 is shown in Figure 7. As in the above steps, it is easy to see that this is a refinement. Since the features are not conflicting, the order in which the blocking and forwarding features are added does not matter.

For all of the above refinements, we have not specified a precedence among possibly overlapping transitions. This is clearly important in some cases, but not is addressed here. It could be achieved by a simple syntactic device, such as adding an ordering on the overlapping transitions. Such an ordering can however easily be translated by adding negations of other conditions, in order to exclude certain transitions. This is a simple refinement step, which we do not model explicitly here. It is a different problem to organize such orderings (i.e. refinements) independent of the features themselves. This important issue is not the goal of this work and is for instance addressed on the programming level in [21].

## 7 Related Work

The state machines proposed in this paper are inspired from I/O-Automata [20]. In contrast to I/O-Automata, where each transition is either labeled by an input message or by an output message, our transitions are labeled by a single input and by a sequence of output messages. Executions of transitions are therefore not regarded to be instantaneous, but they take time to complete. This difference leads to a more compact notation compared to I/O-automata, because intermediate states are not necessary. Moreover, we do not have to regard any further fairness constraints as this is the case in the I/O-automata approach. This is an important prerequisite for the correctness of our refinement calculus. A more detailed discussion of this topic can be found in [22].

Another related approach are Actors [2], which are also automata communicating asynchronously by the exchange of messages. However, Actors are only deterministic and therefore more a programming language than a specification technique. In particular, it is not possible to use them for the abstract specification of component behavior and to define a refinement calculus similar to ours for them.

Several authors have proposed formal approaches for the detection of feature interactions. In [6], an automata-theoretic approach for the detection of feature interactions is proposed. The approach requires a "specific specification style", where each transition of an automaton corresponds to exactly one feature, and where transitions are incrementally added to an existing automaton. Based on this specification style, criteria and rules for interaction detection and resolution are given. However, in our opinion important semantic questions remain unsolved. For instance, due to the lack of a formal semantic model of an information processing system, it is not clear what the underlying notion of "correctness" for the proposed development steps is.

The approach in [3] is based on the temporal logic TLA [19], for which a variety of well-developed specification and refinement techniques exist [1]. From a semantic point of view, the important difference between this approach and ours is that they explicitly abstract from a particular model of communication in order to specify a high-level view of services. Our example has shown that such an abstraction is not necessary, but that a system model based on an asynchronous communication can contribute to the goal of a stepwise process of incrementally adding features. Another important difference is the used notation. Although at first sight this aspect may only seem to be a matter of taste, the difference in complexity of the formulas in [3] and our graphical specifications is striking. The difference may be even greater if one compares the effort we needed to prove that features are compatible with [3]. Unfortunately this question can not be answered. The reason is that although it is formally defined what it means that two features are compatible, no formal proofs showing the compatibility of features are contained in this paper.

## 8 Conclusions

We have presented a graphic description technique using state transition diagrams, which enjoys a formal semantics based on sets of stream processing functions. The theory of stream processing functions provides a simple refinement relation by using set inclusion. This refinement relation is carried over to state transition diagrams as a calculus of formal refinement steps, that can be expressed within our graphic notation.

We have applied this graphical description technique to component and feature specifications. The refinement calculus can nicely be used to add features to an existing component specification. The concept of under-specification has proved to be essential for this approach. If there exists a common refinement, two features can be composed in this way, otherwise, the two features are conflicting and cannot be resolved.

## References

[1] M. Abadi and L. Lamport. The Existence of Refinement Mappings. SRC Research Report 29, Digital Equipment Corporation, 1988.

[2] G.A. Agha. *ACTORS: A Model of Concurrent Computation in Distributed Systems*. MIT-Press, Cambridge, Massachusetts, 1986.

[3] J. Blom, B. Jonsson, and L. Kempe. Using temporal logic for modular specification of telephone services. In Bouma and Velthuijsen [4], pages 197–216.

[4] L. G. Bouma and Hugo Velthuijsen, editors. *Feature Interactions in Telecommunications Systems*. IOS Press, Amsterdam, 1994.

[5] R. Braek and Ø. Haugen. *Engineering Real Time Systems: An object-oriented methodology using SDL*. Prentice Hall, Inc., Englewood Cliffs, New Jersey, 1993.

[6] Jan Bredereke. Formal criteria for feature interactions in telecommunications systems. In *IFIP International Working Conference on Intelligent Networks, Proceedings*, pages 83–97, 1995.

[7] M. Broy. Towards a Formal Foundation of the Specification and Description Language SDL. *Formal Aspects of Computing*, (3):21–57, 1991.

[8] M. Broy. Compositional Refinement of Interactive Systems. Technical Report 89, Digital Equipment Corporation, Systems Research Center, Palo Alto, California, July 1992.

[9] M. Broy, F. Dederichs, C. Dendorfer, M. Fuchs, T.F. Gritzner, and R. Weber. The Design of Distributed Systems - An Introduction to FOCUS. Technical Report SFB 342/2/92 A, Technische Universität München, Institut für Informatik, 1993.

[10] M. Broy and K. Stølen. *Interactive System Design*. Springer-Verlag, 1997. To appear.

[11] E.J. Cameron, N. Griffeth, Y.-J. Linand M.E. Nilson, W.K. Schnure, and H. Velthuijsen. A feature-interaction benchmark for IN and beyond. *IEEE Communications Magazine*, 31(3):64–69, March 1993.

[12] K. E. Cheng and T. Ohta, editors. *Feature Interactions in Telecommunications III*. IOS Press, Tokyo, Japan, Oct 1995.

[13] Radu Grosu, Cornel Klein, and Bernhard Rumpe. Enhancing the SysLab System Model with State . TUM-I 9631, Technische Universität München, 1996.

[14] Radu Grosu, Cornel Klein, Bernhard Rumpe, and Manfred Broy. State transition diagrams. TUM-I 9630, Technische Universität München, 1996.

[15] D. Harel. Statecharts: A visual formalism for complex systems. *Science of Computer Programming*, 8, 1987.

[16] G. Kahn. The semantics of a simple language for parallel programming. *Information Processing*, 74:471–475, 1974.

[17] C. Klein. Prototyping-orientierte Anforderungsspezifikation. Phd thesis, in preparation, 1997.

[18] C. Klein, B. Rumpe, and M. Broy. A Stream based Mathematical Model for Distributed Information Processing Systems. In Elie Najm, editor, *1st Workshop on Formal Methods for Open Object-based Distributed Systems, Paris 1996. Proceedings*. Chapmann & Hall, 1996.

[19] L. Lamport. The temporal logic of actions. Technical Report 79, Digital Equipment Corporation, Systems Research Center, Palo Alto, California, December 1991.

[20] Nancy Lynch and Mark Tuttle. An introduction to Input/Output automata. *CWI Quarterly*, 2(3):219–246, 1989.

[21] Christian Prehofer. An object-oriented approach to feature interaction. In *this volume*, 1997.

[22] B. Rumpe. *Formale Methodik für den Entwurf verteilter objektorientierter Systeme*. PhD thesis, Technische Universität München, 1996.

[23] B. Rumpe and C. Klein. Automata describing object behavior. In H. Kilov and W. Harvey, editors, *Specification of Behavioral Semantics in Object-Oriented Information Modeling*, pages 265–286, Norwell, Massachusetts, 1996. Kluwer Academic Publishers.

[24] P. Zave. Feature interactions and formal specifications in telecommunications. *IEEE Computer*, XXVI(8), August 1993.

[25] P. Zave. Secrets of call forwarding: A specification case study. In *Formal Description Techniques VIII (Proceedings of the Eighth International IFIP Conference on Formal Description Techniques for Distributed Systems and Communication Protocols)*, pages 153–168. Chapmann & Hall, 1996.